# Spin Waves in Quantum Antiferromagnets

B. Kleine, G.S. Uhrig and E. Müller-Hartmann

*Institut für Theoretische Physik, Universität zu Köln D-50937 Köln*

*Phone: 49/221/4704208; Fax: 49/221/4705159*

## Abstract

Using a self-consistent mean-field theory for the $S = 1/2$ Heisenberg antiferromagnet Krüger and Schuck recently derived an analytic expression for the dispersion. It is exact in one dimension ($d = 1$) and agrees well with numerical results in $d = 2$. With an expansion in powers of the inverse coordination number $1/Z$ ($Z = 2d$) we investigate if this expression can be *exact* for all $d$. The projection method of Mori-Zwanzig is used for the *dynamical* spin susceptibility. We find that the expression of Krüger and Schuck deviates in order $1/Z^2$ from our rigorous result. Our method is generalised to arbitrary spin $S$ and to models with easy-axis anisotropy $\Delta$. It can be systematically improved to higher orders in $1/Z$. We clarify its relation to the $1/S$ expansion.

PACS: 75.10, 75.10J, 75.30D

The renewed interest in the Heisenberg quantum antiferrromagnet which was raised by the discovery of the high $T_c$ superconductors and their undoped antiferromagnetic mother substances such as $La_2CuO_4$ is still very vivid [1–7] (for a review see ref. [8]). There are very few exact facts known about the Heisenberg antiferrromagnet except in $d = 1$ where it can be solved by Bethe ansatz. It is proven, for instance, that the translational symmetry is broken in the ground state for all spins $S$, anisotropies $\Delta$ and dimensions $d \geq 2$ except for $d = 2$, $S = 1/2$ and $|\Delta| \in [0.68, 4.55]$ [9].

Besides static quantities such as the staggered magnetization the dynamical quantities such as the frequency-dependent spin susceptibility $\chi(\omega, \mathbf{k})$ are of interest. The frequencies at which the latter diverges define the spin wave dispersion relation $\omega(\mathbf{k})$. Furthermore, the transverse susceptibility and the response of neutron scattering experiments can be deduced from $\chi(\omega, \mathbf{k})$. In the traditional analytic approaches for calculating $\chi(\omega, \mathbf{k})$ which rely on an expansion in $1/S$ one does not evaluate $\chi(\omega, \mathbf{k})$ directly but a bosonic Green function. This results either from a Dyson-Maleev transformation (see e.g. [4]) or from a Holstein-Primakoff transformation (see e.g. [5]).

Very recently, Krüger and Schuck proposed a mean-field theory for the spin wave dispersion of the $S = 1/2$ Heisenberg antiferromagnet on hypercubic lattices in $d$ dimensions by using a Dyson equation approach to calculate the dynamical spin susceptibility [7]. They found a disperson coinciding with the dispersion known from linear spin-wave theory modified by a renormalisation of the spin wave velocity $Z_c$ given by the expression

$$Z_c|_{\text{KS}} = \left( \frac{2}{N} \sum_{\text{MBZ}} \sqrt{1 - \gamma^2(\mathbf{k})} \right)^{-1} \quad (\approx 1.188 \text{ for } Z = 4). \tag{1}$$

Here $\gamma(\mathbf{k}) = (1/d) \sum_{i=1}^{d} \cos(k_i)$ and MBZ stands for the magnetic Brillouin zone ($\gamma(\mathbf{k}) > 0$). In $d = 1$ and for system size $N \to \infty$ (1) yields the correct value of $Z_c = \pi/2$. For $d = 2$ the value of (1), $Z_c|_{\text{KS}} = 1.188$, compares very well, too, with other results obtained by series expansion ($1.18 \pm 0.02$) [1] or $1/S$ expansion ($1.1765 \pm 0.0002$) [4] and ($1.1794$) [5] or by quantum Monte-Carlo methods ($1.14 \pm 0.05$) [2] and ($1.21 \pm 0.03$) [6]. Moreover, we observe that the expansion of the r.h.s. of (1) in powers of the inverse coordination number $1/Z$,

$$Z_c|_{\text{KS}} = 1 + \frac{1}{2Z} + \frac{5}{8Z^2} + \mathcal{O}\left(\frac{1}{Z^3}\right) \quad (\approx 1.164 \text{ for } Z = 4), \tag{2}$$

is correct as far as the rigorously known coefficients of the first two terms (orders 1 and $1/Z$) are concerned [8].

These facts suggest to pose the question if (1) is rigorously valid in all dimensions. In the present Letter we will calculate the coefficient of $1/Z^2$ rigorously by an appropriate application of the projection method of Mori-Zwanzig. We will show that the coefficient 5/8 in (2) is not correct.

We will use the following appropriately scaled Hamiltonian

$$H = \frac{1}{SZ} \sum_{<a,b>} \left( S_a^z S_b^z + \frac{\Delta}{2}(S_a^+ S_b^- + S_a^- S_b^+) \right) \qquad (3)$$

on a $d$-dimensional hypercubic lattice. The sum $<a,b>$ runs over adjacent pairs of sites $a \in A$ and $b \in B$ on the two sublattices $A$ and $B$. The size of the spins is $S$; the coordination number is $Z = 2d$; the anisotropy parameter $\Delta$ controls the spin fluctuations. We will present our approach for the isotropic case $|\Delta| = 1$ to avoid that the main ideas are obscured by calculational details, but will state the results for general $|\Delta| \in [0, 1]$.

The projection method of Mori and Zwanzig [10] is designed to compute resolvents, i.e. dynamical quantities, as continued fraction expansions in terms of static correlations. We employ essentially the notation of Fulde [11]. An operator set $\{A_i\}$ and an antilinear operator product $(A|B)$ are chosen. Here the commutator $(A|B) := \langle 0|[A^+, B]|0\rangle$ with $|0\rangle$ being the ground state is the appropriate choice [12]. Note that this is not a positive semi-definite scalar product. The minimum operator set we use is $A = \{A_1, A_2\}$ with the spin-flip operators $A_1(\mathbf{k}) = \sqrt{2/N} \sum_{i \in A} S_i^- \exp(i\mathbf{k}\mathbf{R}_i)$ and $A_2(\mathbf{k}) = \sqrt{2/N} \sum_{j \in B} S_j^- \exp(i\mathbf{k}\mathbf{R}_j)$.

The Mori-Zwanzig method is applied to the time-dependent spin-spin correlation function $R_{i,j}(t) := -i\langle \mathcal{T}\{ A_i^+(t)A_j\}\rangle$, where $\mathcal{T}$ is the time ordering operator. The causal $2 \times 2$ matrix Green function $R_{i,j}(t)$ contains all information about $\chi(\omega, \mathbf{k})$. With the Liouville operator $L := [H, \ ]$ the Fourier transform $R_{i,j}(\omega)$ equals $R_{i,j}(\omega) = \left(A_i \middle| [\omega - L]^{-1} A_j\right)$. The resolvent $\mathbf{R}(\omega)$ can be rewritten [11]

$$\mathbf{R}(\omega) = \left[\omega - (\mathbf{L} + \mathbf{M}(\omega))\mathbf{P}^{-1}\right]^{-1} \mathbf{P}, \qquad (4)$$

where bold face letters stand for matrices. The matrix elements of $\mathbf{P}, \mathbf{L}$ and $\mathbf{M}(\omega)$ are defined by

$$P_{i,j} := (A_i|A_j), \quad L_{i,j} := (A_i|LA_j), \quad M_{i,j}(\omega) := \left(Q_A L A_i \middle| [\omega - Q_A L Q_A]^{-1} Q_A L A_j\right), \qquad (5)$$

where $Q_A$ is the projection operator onto the operator subspace orthogonal to $A = \{A_1, A_2\}$, i.e. $Q_A := 1 - \sum_{i,j} |A_i) (\mathbf{P}^{-1})_{i,j} (A_j|$.

One obtains an approximation for $\mathbf{R}(\omega)$ by neglecting the memory matrix $\mathbf{M}(\omega)$. This approximation is improved by including more and more operators in the operator set considered [11]. The basic Lanczos idea that new operators are generated by the iterated application of the Liouville operator $L$ serves as a first guideline to which operators should be included. The spin wave dispersion $\omega(\mathbf{k})$ is determined from the poles of the approximate expression for $\mathbf{R}(\omega)$.

The memory matrix $\mathbf{M}(\omega)$ is of the same form as $\mathbf{R}(\omega)$. With $A_i' := Q_A L A_i$ and $L' := Q_A L Q_A$ one can define matrices $\mathbf{P}'$, $\mathbf{L}'$ and $\mathbf{M}'(\omega)$ analogously to (5) and express $\mathbf{M}(\omega)$ according to (4). If we are able to calculate the order in $1/Z$ of the memory matrix $\mathbf{M}(\omega)$ we know in which order an approximate expression for $\mathbf{R}(\omega)$ neglecting $\mathbf{M}(\omega)$ is exact. In practice, we will compute the iterated projector matrix $\mathbf{P}'$ of $\mathbf{M}$ with $P_{i,j}' := (Q_A L A_i | Q_A L A_j)$ and determine its order in $1/Z$. This order obviously is the order of $\mathbf{M}(\omega)$ for large $\omega$. It is also the order of the error of the spin wave dispersion $\omega(\mathbf{k})$, which is determined from the approximation for $\mathbf{R}(\omega)$, because $\mathbf{M}(\omega)$ does not have poles around $\omega = \omega(\mathbf{k})$ [4].

To illustrate the procedure described above we calculate for the set $A$

$$\mathbf{P}_A = 2\langle S_a^z \rangle \begin{bmatrix} 1 & 0 \\ 0 & -1 \end{bmatrix} \quad \mathbf{L}_A = 2\nu \begin{bmatrix} 1 & -\gamma(\mathbf{k}) \\ -\gamma(\mathbf{k}) & 1 \end{bmatrix} \quad \mathbf{P}_A' = \kappa \begin{bmatrix} 1 & 0 \\ 0 & -1 \end{bmatrix}, \quad (6)$$

where $\nu := -(2\langle S_a^z S_b^z \rangle + \langle S_a^+ S_b^- \rangle)/(2S)$ and $a \in A$ and $b \in B$ represent adjacent sites on the two sublattices. The $1/Z$ expansion of the expectation values results in $\langle S_a^z \rangle = S - 1/(4Z) - 9/(16Z^2)$, $\nu = S + (3-6S)/(16SZ^2)$ and $\kappa = -(1-\gamma^2(\mathbf{k}))/Z^2$. From $\kappa = \mathcal{O}(1/Z^2)$ we know that we may trust the dispersion relation resulting from the approximation for $\mathbf{R}(\omega)$ based on $\{A_1, A_2\}$ only to order $1/Z$. In this order one obtains $\omega(\mathbf{k}) = (1 + 1/(4SZ))\sqrt{1 - \gamma^2(\mathbf{k})}$.

To derive (1), Krüger and Schuck inserted (6) into (4), neglected $\mathbf{M}(\omega)$, and used certain self-consistencies specific for $S = 1/2$ to determine the expectation values $\nu$ and $\langle S_a^z \rangle$ [7]. Thus one understands why (1) yields the coefficients up to order $1/Z$ correctly.

The $1/Z$ expansions of the static expectation values needed in the projection method can be computed by ordinary perturbation theory in the spin fluctuations starting from one of the Néel states. To be exact up to order $1/Z^m$ one calculates the perturbation series in $\Delta$ to order $\Delta^{2m}$. Furthermore, the $1/Z$ expansion simplifies complicated integrals over the Brillouin-zone. Any expression containing $\gamma$ is expanded in powers of $\gamma$ which can then be easily summed. In the present calculations we performed these expansions using an algebraic computer programme (MAPLE).

In an attempt to describe the dynamics properly up to order $1/Z^2$ what comes into mind first is the simple Lanczos extension $A' = \{A_1, A_2, A_1', A_2'\}$ of the operator set $\{A_1, A_2\}$ with $A_1' = C \sum_{\langle a,b \rangle} e^{i\mathbf{k}\mathbf{R}_a}(S_b^- S_a^z - S_a^- S_b^z)$ and $A_2' = C \sum_{\langle a,b \rangle} e^{i\mathbf{k}\mathbf{R}_b}(-S_b^- S_a^z + S_a^- S_b^z)$, where $C := Q_A \sqrt{2/N}/(SZ)$. Inspection reveals, however, that the corresponding memory matrix is still of order $1/Z^2$ and not $\mathcal{O}(1/Z^3)$ as desired. Thus this straightforward extension is not sufficient.

Let us consider a general operator set $K$. To describe the dynamics in order $1/Z^2$ correctly it is necessary that for any operator $O \in K$ the deviation of $LO$ from $K$ is $\mathcal{O}(1/Z^3)$. Splitting the Hamiltonian $H = H_{\mathrm{MF}} + H_1$ in $H_{\mathrm{MF}} := -\sum_{a \in A} S_a^z + \sum_{b \in B} S_b^z + NS/2$ and $H_1 := (SZ)^{-1} \sum_{<a,b>} \left[ (S_a^z - S)(S_b^z + S) + (\Delta/2)(S_a^+ S_b^- + S_a^- S_b^+) \right]$ allows to distinghuish between the part of order 1 ($H_{\mathrm{MF}}$) and the part of order $1/Z$ ($H_1$) since the ground state approaches the Néel state on $Z \to \infty$. Commutation with $H_{\mathrm{MF}}$ does not increase the order in powers of $1/Z$. Therefore, we require that $K$ is closed under the Liouville operator $L_{\mathrm{MF}} = [H_{\mathrm{MF}}, \ ]$. Inspecting $A_1'$ and $A_2'$ one realizes that they are not reproduced by $L_{\mathrm{MF}}$ because the relative signs of the two terms on the r.h.s. are changed. This leads us to the set $B = \{B_1, \ldots, B_6\}$ with $B_1 := A_1$, $B_2 := A_2$ and

$$B_3 := -C \sum_{\langle a,b \rangle} S_a^- S_b^z e^{i\mathbf{k}\mathbf{R}_a} \qquad B_4 := C \sum_{\langle a,b \rangle} S_a^- S_b^z e^{i\mathbf{k}\mathbf{R}_b} \tag{7a}$$

$$B_5 := -C \sum_{\langle a,b \rangle} S_b^- S_a^z e^{i\mathbf{k}\mathbf{R}_b} \qquad B_6 := C \sum_{\langle a,b \rangle} S_b^- S_a^z e^{i\mathbf{k}\mathbf{R}_a} \ . \tag{7b}$$

The operators $B_3, \ldots, B_6$ are the split parts of $A_1'$ and $A_2'$: $A_1' = B_3 + B_6$ and $A_2' = B_4 + B_5$. They are eigenoperators of $L_{\mathrm{MF}}$, $L_{\mathrm{MF}} B_i = \pm B_i$, whence $Q_B L_{\mathrm{MF}} B_i = 0$.

Explicit calculation to order $1/Z^2$ yields for the operator set $B$

$$\mathbf{P}_B = \begin{bmatrix} \mathbf{P} & 0 & 0 \\ 0 & \mathbf{N} & 0 \\ 0 & 0 & -\mathbf{N} \end{bmatrix} \quad \mathbf{L}_B = \begin{bmatrix} \mathbf{L} & \mathbf{N} & \mathbf{VN} \\ \mathbf{N} & 3\mathbf{N} & 0 \\ \mathbf{NV} & 0 & 3\mathbf{N} \end{bmatrix} \quad \mathbf{N} := \frac{1}{2SZ^2} \begin{bmatrix} 1+\gamma^2 & -2\gamma \\ -2\gamma & 2 \end{bmatrix}, \tag{8}$$

where $\gamma$ is short for $\gamma(\mathbf{k})$ and the matrix $\mathbf{V} := -[[0, 1], [1, 0]]$. We find a block structure with $2 \times 2$ blocks. The upper left block represents the information that we had already with the first set $\{A_1, A_2\}$. The matrix elements of $\mathbf{P}_B$ which link $B_1, B_2$ with the other operators are zero by construction since $Q_A B_{1/2} = 0$. The corresponding lateral matrix elements of $\mathbf{L}_B$ are easily obtained from $(B_i|LB_1) = (B_i|B_3 + B_6)$ and $(B_i|LB_2) = (B_i|B_4 + B_5)$ for $i > 2$. We expect the simple block structure of the lower right $4 \times 4$-block to disappear in order $1/Z^3$.

Explicit evaluation aided by computer algebra shows that $\mathbf{P}_B'$ of $\mathbf{M}_B$ is $\mathcal{O}(1/Z^3)$. By the line of argument presented above we know therefore that the dynamics is now correctly described to order $1/Z^2$.

Using (8) and neglecting $\mathbf{M}_B(\omega)$ we obtain a $6 \times 6$-resolvent. From this we use the upper left $2 \times 2$-block as improved approximation for $\mathbf{R}(\omega)$ which is correct to order $1/Z^2$. The zeros of $\det\left[\omega - \mathbf{L}_B \mathbf{P}_B^{-1}\right]$ give the dispersion relation $\omega(\mathbf{k})$. One finds $\det\left[\omega - \mathbf{L}_B \mathbf{P}_B^{-1}\right] = (\omega^2 - 9 + \mathcal{O}(1/Z))^2 (\omega^2 - Z_c^2(1 - \gamma^2(\mathbf{k})) + \mathcal{O}(1/Z^3))$ with

$$Z_c = 1 + \frac{1}{4SZ} + \frac{3}{16SZ^2} + \mathcal{O}\left(\frac{1}{Z^3}\right) \quad (\approx 1.148 \text{ for } Z = 4, S = 1/2). \tag{9}$$

With this renormalisation factor $\omega(\mathbf{k}) = Z_c\sqrt{1-\gamma^2(\mathbf{k})}$ represents the systematic second order $1/Z$ expansion of the spin wave dispersion. To this order the spin fluctuations only renormalise the spin wave velocity; they do not affect the shape of the dispersion. We do, however, expect a modification of the shape from higher orders of the $1/Z$ expansion.

For $S = 1/2$, (9) differs from the Krüger/Schuck result in (2). We therefore come to the conclusion that the analytic expression (1) cannot be rigorously valid in *all* dimensions, not excluding completely the possible validity in any *particular* dimension. There is no fundamental reason known why the mean-field theory of Krüger/Schuck yields the exact results in $d = 1$ although a similar agreement was already found before [13].

Investigating the pole structure given by $\det\left[\omega - \mathbf{L}_B\mathbf{P}_B^{-1}\right]$ one finds that besides the dispersion pole there is a *double* pole at $\omega = 3$ which is dispersionless. Physically it corresponds to 3-magnon processes. They are dispersionless since we desribe the 3-magnon-pole only in leading order where due to $\sqrt{1-\gamma^2} = 1 + \mathcal{O}(1/Z)$ on averaging in the Brillouin zone the magnon energies are replaced by 1.

It is obvious that the approach described above can be extended to the model (3) with an easy-axis anisotropy $|\Delta| \leq 1$. We state here the final second order result for the dispersion relation

$$\omega^2(\mathbf{k}) = 1 - \Delta^2\gamma^2 + \frac{\Delta^2(1+\Delta^2\gamma^2-2\gamma^2)}{2SZ} + \frac{3\Delta^4(3\Delta^2\gamma^2-4\gamma^2+1)}{8SZ^2} - \frac{\Delta^2\left(9\Delta^6\gamma^4 - 3\Delta^4\gamma^2(4\gamma^2-29) + 4\Delta^2(\gamma^4-52\gamma^2+6) + 128\gamma^2 - 32\right)}{16S^2Z^2(8+\Delta^2\gamma^2)} + \mathcal{O}\left(\frac{1}{Z^3}\right) . \quad (10)$$

There is an interesting relation between the $1/S$ expansion and the $1/Z$ expansion of the spin wave dispersion: the correction of order $S^{-m}$ in the $1/S$ expansion calculated for a hypercubic lattice of arbitrary dimension is small of order $Z^{-m}$. This implies that taking this $1/S$ expansion up to the order $S^{-m}$ one obtains all contributions up to the order $Z^{-m}$ of the $1/Z$ expansion. This also explains why in our $1/Z$ expansion terms of order $Z^{-m}$ contain only contributions of order $S^{-k}$ with $k \leq m$.

Note that the hydrodynamic relation $Z_c^2 = Z_\rho/Z_\chi$ [8] is intrinsically fulfilled since *all* three renormalisations can be derived from $\mathbf{R}(\omega)$. With the dynamic transverse susceptibility $\chi(\omega, \mathbf{k}) := -(R_{1,1} + 2R_{1,2} + R_{2,2})/4$ one finds
$Z_\chi = \lim_{\mathbf{k}\to 0}\lim_{\omega\to 0}(1/\chi_\text{class})\chi(\omega, \mathbf{k})$ and for the spin stiffness renormalisation $Z_\rho = -\lim_{\omega\to 0}\lim_{\mathbf{k}\to 0}(1/\rho_\text{class})(\omega^2/\mathbf{k}^2)\chi(\omega, \mathbf{k})$ [14]. Our results are $Z_\chi = 1 - 1/(2SZ) + (5 - 18S)/(24S^2Z^2) + \mathcal{O}(Z^{-3})$ and $Z_\rho = 1 + (1-18S)/(48S^2Z^2) + \mathcal{O}(Z^{-3})$.

In conclusion, a systematic $1/Z$ expansion of the dynamical spin correlation function $\mathbf{R}(\omega)$ up to the order $1/Z^2$ was presented. Explicit results for arbitrary spin $S$ and easy-axis anisotropies $|\Delta| \leq 1$ for the spin wave dispersion up to the order $1/Z^2$ were given. The result for $S = 1/2$ and $|\Delta| = 1$ was used to conclude that the spin wave velocity (1) of Krüger and Schuck is not exact in *all* dimensions. A general recipe how to extend the $1/Z$ expansion to higher orders was given: One has to make sure that the set of operators generated by iterated application of the Liouville operator $L$ is closed under $L_{\mathrm{MF}}$. Finally, the relation of the $1/Z$ expansion to the $1/S$ expansion was elucidated.

Acknowledgments

We gratefully acknowledge helpful discussions with K. Becker, W. Brenig and A. Kampf. This work was supported by the Deutsche Forschungsgemeinschaft in the SFB 341.

# References

[1] R. R. P. Singh, Phys. Rev. B **39**, 9760 (1989); R. R. P. Singh and D. A. Huse, Phys. Rev. B **40**, 7247 (1989).

[2] N. Trivedi and D. M. Ceperley, Phys. Rev. B **40**, 2737 (1989)

[3] Z. Weihong, J. Oitmaa and C. J. Hamer, Phys. Rev. B **43**, 8321 (1991).

[4] C. M. Canali, S. M. Girvin and M. Wallin, Phys. Rev. B **45**, 10131 (1992); C. M. Canali and S. M. Girvin, Phys. Rev. B **45**, 7127 (1992),

[5] J. Igarashi, Phys. Rev. Lett. **46**, 10763 (1992).

[6] G. Chen, H.-Q. Ding and W. A. Goddard III, Phys. Rev. B **46**, 2933 (1992)

[7] P. Krüger and P. Schuck, Europhys. Lett. **27**, 395 (1994).

[8] E. Manousakis, Rev. Mod. Phys. **63**, 1 (1991)

[9] H.-A. Wischmann and E. Müller-Hartmann, Phys. Rev. B **43**, 8668 (1991); J. Phys. I **1**, 647 (1991) and refs. therein

[10] H. Mori, Prog. Theor. Phys. **33**, 423 (1965); R. Zwanzig, in *Lectures in Theoretical Physics*, edited by W. E. Brittin, B. W. Downs and J. Downs, vol. III (Interscience, New York, 1961)

[11] P. Fulde, *Electron Correlations in Molecules and Solids*, vol. 100 of *Solid State Sciences* (Springer-Verlag, Berlin, 1993), appendix C

[12] K. W. Becker and U. Muschelknautz, Phys. Rev. B **48**, 13826 (1993)

[13] A. Jevicki and N. Papanicolaou, Ann. Phys. (NY) **120**, 107 (1979)

[14] In order that $Z_\chi$ and $Z_\rho$ exist, $\chi(\omega, \mathbf{k})$ has the form $a\mathbf{k}^2/(c^2\mathbf{k}^2 - \omega^2)$ for small values of $\omega$ and $\mathbf{k}$. The variables $c$ and $a$ are non-vanishing constants for $\omega \to 0, \mathbf{k} \to 0$, $c$ being the spin wave velocity. This form implies the hydrodynamic relation.